# Hybrid-Logical Reasoning in False-Belief Tasks

Torben Braüner
Programming, Logic and Intelligent Systems Research Group
Roskilde University
P.O. Box 260
DK-4000 Roskilde, Denmark
torben@ruc.dk

## ABSTRACT

The main aim of the present paper is to use a proof system for hybrid modal logic to formalize what are called false-belief tasks in cognitive psychology, thereby investigating the interplay between cognition and logical reasoning about belief. We consider two different versions of the Smarties task, involving respectively a shift of perspective to another person and to another time. Our formalizations disclose that despite this difference, the two versions of the Smarties task have exactly the same underlying logical structure. We also consider the Sally-Anne task, having a somewhat more complicated logical structure, presupposing a "principle of inertia" saying that a belief is preserved over time, unless there is belief to the contrary.

## 1. INTRODUCTION

In the area of cognitive psychology there is a reasoning task called the *Smarties task*. The following is one version of the Smarties task.

> A child is shown a Smarties tube where unbeknownst to the child the Smarties have been replaced by pencils. The child is asked: "What do you think is inside the tube?" The child answers "Smarties!" The tube is then shown to contain pencils only. The child is then asked: "If your mother comes into the room and we show this tube to her, what will she think is inside?"

It is well-known from experiments that most childred above the age of four correctly say "Smarties" (thereby attributing a false belief to the mother) whereas younger children say "Pencils" (what they know is inside the tube). For autistic[1] children the cutoff age is higher than four years, which is one reason to the interest in the Smarties task.

The Smarties task is one out of a family of reasoning tasks called *false-belief tasks* showing the same pattern, that most children above four answer correctly, but autistic children have to be older. This was first observed in the paper [4] in connection with another false-belief task called the *Sally-Anne task*. Starting with the authors of that paper, many researchers in cognitive psychology have argued that there is a link between autism and a lack of what is called *theory of mind*, which is a capacity to imagine other people's mental states, for example their beliefs. For a very general formulation of the theory of mind deficit hypothesis of autism, see the book [3].

---

[1]Autism is a psychiatric disorder with the following three diagnostic criteria: 1. Impairment in social interaction. 2. Impairment in communication. 3. Restricted repetitive and stereotyped patterns of behavior, interests, and activities. For details, see *Diagnostic and Statistical Manual of Mental Disorders, 4th Edition (DSM-IV)*, published by the American Psychiatric Association.

Giving a correct answer to the Smarties task involves a shift of perspective to another person, namely the mother. You have to put yourself in another person's shoes, so to speak. Since the capacity to take another perspective is a precondition for figuring out the correct answer to the Smarties task and other false-belief tasks, the fact that autistic children have a higher cutoff age is taken to support the claim that autists have a limited or delayed theory of mind. For a critical overview of these arguments, see the book [23] by Keith Stenning and Michiel van Lambalgen. The books [23] and [3] not only consider theory of mind at a cognitive level, such as in connection with false-belief tasks, but they also discuss it from a biological point of view.

In a range of works van Lambalgen and co-authors have given a detailed logical analysis (but not a full formalization) of the reasoning taking place in the Smarties task and other false-belief tasks in terms of closed-world reasoning as used in non-monotonic logics, see in particular [23]. The analysis of the Smarties task of [23] (in Subsection 9.4.4) makes use of a modality $B$ for belief satisfying two standard modal principles.[2] The first principle is $B(\phi \to \psi) \to (B\phi \to B\psi)$ (principle (9.5) at page 251 in [23]). The second principle is the rule called necessitation, that is, from $\phi$ derive $B\phi$ (this principle is not mentioned explicitly in [23], but is implicit in the analysis given at the bottom of page 256). These two principles together imply that belief is closed under logical consequence, that is, $B\psi$ can be derived from $\phi \to \psi$ together with $B\phi$, which at least for human agents is implausible (when the modal operator stands for knowledge, this is called logical omniscience).

In the present paper we give a logical analysis of the perspective shift required to give a correct answer to the Smarties and Sally-Anne tasks, and we demonstrate that these tasks can be fully formalized in a hybrid-logical proof system not assuming principles implying logical omniscience, namely the natural deduction system described in Chapter

---

[2]Strictly speaking, the modality $B$ in [23] is not formalized in terms of modal logic, but in terms of what is called event calculus, where $B$ is a predicate that can take formulas as arguments.

4 of the book [8], and the paper [7] as well. Beside not suffering from logical omniscience, why is a *natural deduction* system for *hybrid modal logic* appropriate to this end?

- The subject of proof-theory is the notion of proof and formal, that is, symbolic, systems for representing proofs. Formal proofs built according to the rules of proof systems can be used to represent—describe the structure of—mathematical arguments as well as arguments in everyday human practice. Beside giving a way to distinguish logically correct arguments from incorrect ones, proof systems also give a number of ways to characterize the structure of arguments. Natural deduction style proofs are meant to formalize the way human beings actually reason, so natural deduction is an obvious candidate when looking for a proof system to formalize the Smarties task in.

- In the standard Kripke semantics for modal logic, the truth-value of a formula is relative to points in a set, that is, a formula is evaluated "locally" at a point, where points usually are taken to represent possible worlds, times, locations, epistemic states, persons, states in a computer, or something else. Hybrid logics are extended modal logics where it is possible to directly refer to such points in the logical object language, whereby locality can be handled explicitly, for example, when reasoning about time one can formulate a series of statements about what happens at specific times, which is not possible in ordinary modal logic. Thus, when points in the Kripke semantics represent local perspectives, hybrid-logical machinery can handle explicitly the different perspectives in the Smarties task.

For the above reasons, we have been able to turn our informal logical analysis of the Smarties and Sally-Anne tasks into formal hybrid-logical natural deduction proofs closely reflecting the shift between different perspectives.

The natural deduction system we use for our formalizations is a modified version of a natural deduction system for a logic of situations similar to hybrid logic, originally introduced in the paper [19] by Jerry Seligman. The modified system was introduced in the paper [7], and later on considered in Chapter 4 of the book [8], both by the present author. In what follows we shall simply refer to the modified system as Seligman's system.

Now, Seligman's system allows any formula to occur in it, which is different from the most common proof systems for hybrid logic that only allow formulas of a certain form called satisfaction statements. This is related to a different way of reasoning in Seligman's system, which captures particularly well the reasoning in the Smarties and Sally-Anne tasks. We prove a completeness result which also says that Seligman's system is analytic, that is, we prove that any valid formula has a derivation satisfying the subformula property. Analyticity guarentees that any valid argument can be formalized using only subformulas of the premises and the conclusion. The notion of analyticity goes back to G.W. Leibniz (1646–1716) who called a proof analytic if and only if the proof is based on concepts contained in the proven statement, the main aim being to be able to construct a proof by an analysis of the result, cf. [2].

The present paper is structured as follows. In the second section we recapitulate the basics of hybrid logic, readers well-versed in hybrid logic can safely skip this section. In the third section we introduce Seligman's natural deduction system for hybrid logic. In the fourth and fifth sections we formalize two versions of the Smarties task using this system, and in the sixth section we formalize the Sally-Anne task. A discussion can be found in the seventh section, in the eightth section there are some brief remarks on other work, and in the final section some remarks on further work. In the appendix we prove the above mentioned completeness result, which also demonstrates analyticity.

## 2. HYBRID LOGIC

The term "hybrid logic" covers a number of logics obtained by adding further expressive power to ordinary modal logic. The history of what now is known as hybrid logic goes back to the philosopher Arthur Prior's work in the 1960s. See the handbook chapter [1] for a detailed overview of hybrid logic. See the book [8] on hybrid logic and its proof-theory.

The most basic hybrid logic is obtained by extending ordinary modal logic with *nominals*, which are propositional symbols of a new sort. In the Kripke semantics a nominal is interpreted in a restricted way such that it is true at exactly one point. If the points are given a temporal reading, this enables the formalization of natural language statements that are true at exactly one time, for example

> it is five o'clock May 10th 2007

which is true at the time five o'clock May 10th 2007, but false at all other times. Such statements cannot be formalized in ordinary modal logic, the reason being that there is only one sort of propositional symbol available, namely ordinary propositional symbols, which are not restricted to being true at exactly one point.

Most hybrid logics involve further additional machinery than nominals. There is a number of options for adding further machinery; here we shall consider a kind of operator called *satisfaction operators*. The motivation for adding satisfaction operators is to be able to formalize a statement being true at a particular time, possible world, or something else. For example, we want to be able to formalize that the statement "it is raining" is true at the time five o'clock May 10th 2007, that is, that

> at five o'clock May 10th 2007 it is raining.

This is formalized by the formula $@_a r$ where the nominal $a$ stands for "it is five o'clock May 10th 2007" as above and where $r$ is an ordinary propositional symbol that stands for "it is raining". It is the part $@_a$ of the formula $@_a r$ that is called a satisfaction operator. In general, if $a$ is a nominal and $\phi$ is an arbitrary formula, then a new formula $@_a \phi$ can be built (in some literature the notation $a : \phi$ is used instead of $@_a \phi$). A formula of this form is called a *satisfaction statement*. The formula $@_a \phi$ expresses that the formula $\phi$ is true at one particular point, namely the point to which the nominal $a$ refers. Nominals and satisfaction operators are the most common pieces of hybrid-logical machinery, and are what we need for the purpose of the present paper.

In what follows we give the formal syntax and semantics of hybrid logic. It is assumed that a set of ordinary propositional symbols and a countably infinite set of nominals are given. The sets are assumed to be disjoint. The metavariables $p$, $q$, $r$, ... range over ordinary propositional symbols and $a, b, c, \ldots$ range over nominals. Formulas are defined by the following grammar.

$$S ::= p \mid a \mid S \wedge S \mid S \rightarrow S \mid \bot \mid \Box S \mid @_a S$$

The metavariables $\phi, \psi, \theta, \ldots$ range over formulas. Negation is defined by the convention that $\neg \phi$ is an abbreviation for $\phi \rightarrow \bot$. Similarly, $\Diamond \phi$ is an abbreviation for $\neg \Box \neg \phi$.

DEFINITION 2.1. *A* model *for hybrid logic is a tuple*

$$(W, R, \{V_w\}_{w \in W})$$

*where*

1. *$W$ is a non-empty set;*
2. *$R$ is a binary relation on $W$; and*
3. *for each $w$, $V_w$ is a function that to each ordinary propositional symbol assigns an element of $\{0, 1\}$.*

The pair $(W, R)$ is called a *frame*. Note that a model for hybrid logic is the same as a model for ordinary modal logic. Given a model $\mathfrak{M} = (W, R, \{V_w\}_{w \in W})$, an *assignment* is a function $g$ that to each nominal assigns an element of $W$. The relation $\mathfrak{M}, g, w \models \phi$ is defined by induction, where $g$ is an assignment, $w$ is an element of $W$, and $\phi$ is a formula.

$$\begin{array}{rcl}
\mathfrak{M}, g, w \models p & \text{iff} & V_w(p) = 1 \\
\mathfrak{M}, g, w \models a & \text{iff} & w = g(a) \\
\mathfrak{M}, g, w \models \phi \wedge \psi & \text{iff} & \mathfrak{M}, g, w \models \phi \text{ and } \mathfrak{M}, g, w \models \psi \\
\mathfrak{M}, g, w \models \phi \rightarrow \psi & \text{iff} & \mathfrak{M}, g, w \models \phi \text{ implies } \mathfrak{M}, g, w \models \psi \\
\mathfrak{M}, g, w \models \bot & \text{iff} & \text{falsum} \\
\mathfrak{M}, g, w \models \Box \phi & \text{iff} & \text{for any } v \in W \text{ such that } wRv, \\
& & \mathfrak{M}, g, v \models \phi \\
\mathfrak{M}, g, w \models @_a \phi & \text{iff} & \mathfrak{M}, g, g(a) \models \phi
\end{array}$$

By convention $\mathfrak{M}, g \models \phi$ means $\mathfrak{M}, g, w \models \phi$ for every element $w$ of $W$ and $\mathfrak{M} \models \phi$ means $\mathfrak{M}, g \models \phi$ for every assignment $g$. A formula $\phi$ is *valid* if and only if $\mathfrak{M} \models \phi$ for any model $\mathfrak{M}$.

## 3. SELIGMAN'S SYSTEM

In this section we introduce Seligman's natural deduction systems for hybrid logic. Before defining the system, we shall sketch the basics of natural deduction. Natural deduction style derivation rules for ordinary classical first-order logic were originally introduced by Gerhard Gentzen in [11] and later on developed much further by Dag Prawitz in [16, 17]. See [24] for a general introduction to natural deduction systems. With reference to Gentzen's work, Prawitz made the following remarks on the significance of natural deduction.

> ...the essential logical content of intuitive logical operations that can be formulated in the languages considered can be understood as composed of the atomic inferences isolated by Gentzen. It is in this sense that we may understand the terminology *natural* deduction.
>
> Nevertheless, Gentzen's systems are also natural in the more superficial sense of corresponding rather well to informal practices; in other words, the structure of informal proofs are often preserved rather well when formalized within the systems of natural deduction. ([17], p. 245)

Similar views on natural deduction are expressed many places, for example in a textbook by Warren Goldfarb.

> What we shall present is a system for *deductions*, sometimes called a system of *natural deduction*, because to a certain extent it mimics certain natural ways we reason informally. In particular, at any stage in a deduction we may introduce a new premise (that is, a new supposition); we may then infer things from this premise and eventually eliminate the premise (*discharge* it). ([13], p. 181)

Basically, what is said by the second part of the quotation by Prawitz, and the quotation by Goldfarb as well, is that the structure of informal human arguments can be described by natural deduction derivations.

Of course, the observation that natural deduction derivations often can formalize, or mimic, informal reasoning does not itself prove that natural deduction is the mechanism underlying human deductive reasoning, that is, that formal rules in natural deduction style are somehow built into the human cognitive architecture. However, this view is held by a number of psychologists, for example Lance Rips in the book [18], where he provides experimental support for the claim.

> ...a person faced with a task involving deduction attempts to carry it out through a series of steps that takes him or her from an initial description of the problem to its solution. These intermediate steps are licensed by mental inference rules, such as modus ponens, whose output people find intuitively obvious. ([18], p. x)

This is the main claim of the "mental logic" school in the psychology of reasoning (whose major competitor is the "mental models" school, claiming that the mechanism underlying human reasoning is the construction of models, rather than the application of topic-neutral formal rules).

We have now given a brief motivation for natural deduction and proceed to a formal definition. A *derivation* in a natural deduction system has the form of a finite tree where the nodes are labelled with formulas such that for any formula occurrence $\phi$ in the derivation, either $\phi$ is a leaf of the derivation or the immediate successors of $\phi$ in the derivation are the premises of a rule-instance which has $\phi$ as the conclusion. In what follows, the metavariables $\pi, \tau, \ldots$ range over derivations. A formula occurrence that is a leaf is called an *assumption* of the derivation. The root of a derivation is called the *end-formula* of the derivation. All assumptions are annotated with numbers. An assumption is either *undischarged* or *discharged*. If an assumption is discharged, then it is discharged at one particular rule-instance and this is indicated by annotating the assumption and the rule-instance with identical numbers. We shall often omit this information when no confusion can occur. A rule-instance annotated with some number discharges all undischarged assumptions that are above it and are annotated with the number in question, and moreover, are occurrences of a formula determined by the rule-instance.

Two assumptions in a derivation belong to the same *parcel* if they are annotated with the same number and are occurrences of the same formula, and moreover, either are both

**Figure 1: Rules for connectives**

$$\frac{\phi \quad \psi}{\phi \wedge \psi} \,(\wedge I) \qquad \frac{\phi \wedge \psi}{\phi} \,(\wedge E1) \qquad \frac{\phi \wedge \psi}{\psi} \,(\wedge E2)$$

$$\frac{\begin{array}{c}[\phi]\\ \vdots\\ \psi\end{array}}{\phi \to \psi} \,(\to I) \qquad \frac{\phi \to \psi \quad \phi}{\psi} \,(\to E)$$

$$\frac{\begin{array}{c}[\neg \phi]\\ \vdots\\ \bot\end{array}}{\phi} \,(\bot)^*$$

$$\frac{a \quad \phi}{@_a \phi} \,(@I) \qquad \frac{a \quad @_a \phi}{\phi} \,(@E)$$

$$\frac{\begin{array}{c}[\Diamond c]\\ \vdots\\ @_c \phi\end{array}}{\Box \phi} \,(\Box I)^{\dagger} \qquad \frac{\Box \phi \quad \Diamond e}{@_e \phi} \,(\Box E)$$

∗ $\phi$ is a propositional letter.
† $c$ does not occur free in $\Box \phi$ or in any undischarged assumptions other than the specified occurrences of $\Diamond c$.

**Figure 2: Rules for nominals**

$$\frac{\phi_1 \quad \ldots \quad \phi_n \quad \begin{array}{c}[\phi_1]\ldots[\phi_n][a]\\ \vdots\\ \psi\end{array}}{\psi} \,(Term)^* \qquad \frac{\begin{array}{c}[a]\\ \vdots\\ \psi\end{array}}{\psi} \,(Name)^{\dagger}$$

∗ $\phi_1, \ldots, \phi_n$, and $\psi$ are all satisfaction statements and there are no undischarged assumptions in the derivation of $\psi$ besides the specified occurrences of $\phi_1, \ldots, \phi_n$, and $a$.
† $a$ does not occur in $\psi$ or in any undischarged assumptions other than the specified occurrences of $a$.

undischarged or have both been discharged at the same rule-instance. Thus, in this terminology rules discharge parcels. We shall make use of the standard notation

$$\begin{array}{c}[\phi^r]\\ \vdots \pi\\ \psi\end{array}$$

which means a derivation $\pi$ where $\psi$ is the end-formula and $[\phi^r]$ is the parcel consisting of all undischarged assumptions that have the form $\phi^r$.

We shall make use of the following conventions. The metavariables $\Gamma, \Delta, \ldots$ range over sets of formulas. A derivation $\pi$ is called a *derivation of* $\phi$ if the end-formula of $\pi$ is an occurrence of $\phi$, and moreover, $\pi$ is called a *derivation from* $\Gamma$ if each undischarged assumption in $\pi$ is an occurrence of a formula in $\Gamma$ (note that numbers annotating undischarged assumptions are ignored). If there exists a derivation of $\phi$ from $\emptyset$, then we shall simply say that $\phi$ is *derivable*.

A typical feature of natural deduction is that there are two different kinds of rules for each connective; there are rules called introduction rules which introduce a connective (that is, the connective occurs in the conclusion of the rule, but not in the premises) and there are rules called elimination rules which eliminate a connective (the connective occurs in a premiss of the rule, but not in the conclusion). Introduction rules have names in the form $(\ldots I \ldots)$, and similarly, elimination rules have names in the form $(\ldots E \ldots)$.

Now, Seligman's natural deduction system is obtained from the rules given in Figure 1 and Figure 2. We let $\mathbf{N'}_{\mathcal{H}}$ denote the system thus obtained. The system $\mathbf{N'}_{\mathcal{H}}$ is taken from [7] and Chapter 4 of [8] where it is shown to be sound and complete wrt. the formal semantics given in the previous section. As mentioned earlier, this system is a modified version of a system originally introduced in [19]. The system of [19] was modified in [7] and [8] with the aim of obtaining a desirable property called closure under substitution, see Subsection 4.1.1 of [8] for further explanation.

## 4. A FIRST EXAMPLE

The way of reasoning in Seligman's system is different from the way of reasoning in most other proof systems for hybrid logic[3]. In this section we give the first example of reasoning using the (*Term*) rule (displayed in Figure 2).

Beside the (*Term*) rule, the key rules in the example are the rules (@I) and (@E) (displayed in Figure 1), which are the introduction and elimination rules for the satisfaction operator. The rule (@I) formalizes the following informal argument.

> It is Christmas Eve 2011; it is snowing, so at Christmas Eve 2011 it is snowing.

And the rule (@E) formalizes the following.

> It is Christmas Eve 2011; at Christmas Eve 2011 it is snowing, so it is snowing.

The (*Term*) rule enables hypothetical reasoning where reasoning is about what is the case at a specific time, possibly different from the actual time. Consider the following informal argument.

> At May 10th 2007 it is raining; if it is raining it is wet, so at May 10th 2007 it is wet.

The reasoning in this example argument is about what is the case at May 10th 2007. If this argument is made at a specific actual time, the time of evaluation is first shifted from the actual time to a hypothetical time, namely May 10th 2007, then some reasoning is performed involving the premise "if it is raining it is wet", and finally the time of evaluation is shifted back to the actual time. The reader is invited to verify this shift of time by checking that the argument is correct, and note that the reader himself (or herself) imagines being at the time May 10th 2007. Note that the premise "if it is raining it is wet" represents a causal relation holding at all times.

---

[3]We here have in mind natural deduction, Gentzen, and tableau systems for hybrid logic, not axiom systems. Proof systems of the first three types are suitable for actual reasoning, carried out by a human, a computer, or in some other medium. Axiom systems are usually not meant for actual reasoning, but are of a more foundational interest.

Now, in a temporal setting, the side-condition on the rule (*Term*) requiring that all the formulas $\phi_1, \ldots, \phi_n, \psi$ are satisfaction statements (see Figure 2) ensures that these formulas are temporally definite, that is, they have the same truth-value at all times, so the truth-value of these formulas are not affected by a shift of temporal perspective. The rule would not be sound if the formulas were not temporally definite.

We now proceed to the formalization of the above argument about what is the case at May 10th 2007. We make use of the following symbolizations

- $p$  It is raining
- $q$  It is wet
- $a$  May 10th 2007

and we take the formula $p \rightarrow q$ as an axiom since it represents a causal relation between $p$ and $q$ holding at all times (note that we use an axiom since the relation $p \rightarrow q$ holds between the particular propositions $p$ and $q$, we do not use an axiom schema since the relation obviously does not hold between any pair of propositions).[4] Then the argument can be formalized as the right-hand-side derivation in Figure 3.

It is instructive to see how the right-hand-side derivation in Figure 3 is built, so at the left-hand-side of the figure we have displayed the derivation to which the (*Term*) rule is applied, whereby the right-hand-side derivation is obtained. Thus, the application of the (*Term*) rule in the right-hand-side derivation delimits a piece of reasoning taking place at a certain hypothetical time, which is the left-hand-side derivation.

The above example argument is similar to an example given in the paper [19]. The following is a slightly reformulated version.

> In Abu Dabi alcohol is forbidden; if alcohol is forbidden Sake is forbidden, so in Abu Dabi Sake is forbidden.

Thus, the example of [19] involves spatial locations rather than times, and the shift is to a hypothetical place, namely the city of Abu Dabi.

Formally, the shift to a hypothetical point of evaluation effected by the rule (*Term*) can be seen by inspecting the proof that the rule (*Term*) is sound: The world of evaluation is shifted from the actual world to the hypothetical world where the nominal $a$ is true (see Figure 2), then some reasoning is performed involving the delimited subderivation which by induction is assumed to be sound, and finally the world of evaluation is shifted back to the actual world. Soundness of the system $\mathbf{N}'_\mathcal{H}$, including soundness of the rule (*Term*), is proved in Theorem 4.1 in Section 4.3 of [8].

---
[4]One of the anonymous reviewers asked why the premise "if it is raining it is wet" is formalized as $p \rightarrow q$ using classical implication, rather than a form of non-monotonic implication. Like in many cases when classical logic is used to formalize natural language statements, there is an idealization in our choice of classical implication. We think this idealization is justified since our main goal is to formalize the perspective shift involved in the example argument, which we presume is orthogonal to the issue of non-monotonicity. We note in passing that our premise "if it is raining it is wet" corresponds to the premise "if alcohol is forbidden Sake is forbidden" in Seligman's example argument briefly described below, and Seligman also uses classical implication, or to be precise, machinery equivalent to classical implication, [19]. See also Footnote 6.

The rule (*Term*) is very different from other rules in proof systems for hybrid logic, roughly, this rule replaces rules for equational reasoning in other systems, see for example the rules in the natural deduction system given in Section 2.2 of the book [8].

In passing we mention that the way in which the (*Term*) rule delimits a subderivation is similar to the way subderivations are delimited by so-called boxes in linear logic, and more specifically, the way a subderivation is delimited by the introduction rule for the modal operator □ in the natural deduction system for S4 given in [5], making use of explicit substitutions in derivations.

## 5. THE SMARTIES TASK (TEMPORAL SHIFT VERSION)

In this section we will give a formalization which has exactly the same structure as the formalization in the previous section, but which in other respects is quite different. It turns out that a temporal shift like the one just described in the previous section also takes place in the following version of the Smarties task, where instead of a shift of perspective to another person, there is a shift of perspective to another time.[5]

> A child is shown a Smarties tube where unbeknownst to the child the Smarties have been replaced by pencils. The child is asked: "What do you think is inside the tube?" The child answers "Smarties!" The tube is then shown to contain pencils only. The child is then asked: "Before this tube was opened, what did you think was inside?"

See [14] for more on the temporal version of the Smarties task. Below we shall formalize each step in the logical reasoning taking place when giving a correct answer to the task, but before that, we give an informal analysis. Let us call the child Peter. Let $a$ be the time where Peter answers the first question, and let $t$ be the time where he answers the second one. To answer the second question, Peter imagines himself being at the earlier time $a$ where he was asked the first question. At that time he deduced that there were Smarties inside the tube from the fact that it is a Smarties tube. Imagining being at the time $a$, Peter reasons that since he at that time deduced that there were Smarties inside, he must also have come to believe that there were Smarties inside. Therefore, at the time $t$ he concludes that at the earlier time $a$ he believed that there were Smarties inside.

We now proceed to the full formalization. We first extend the language of hybrid logic with two modal operators, $D$ and $B$. We make use of the following symbolizations

- $D$  Peter deduces that ...
- $B$  Peter believes that ...
- $p$  There are Smarties inside the tube
- $a$  The time where the first question is asked

and we take the principle $D\phi \rightarrow B\phi$ as an axiom schema (it holds whatever proposition is substituted for the metavariable $\phi$, hence an axiom schema). This is principle (9.4) in

$$\cfrac{a \quad \cfrac{\cfrac{a \quad @_a p}{p}(@E) \quad \cfrac{}{p \to q}(Axiom)}{q}(\to E)}{@_a q}(@I)$$

$$\cfrac{@_a p \quad \cfrac{[a] \quad \cfrac{\cfrac{[a] \quad [@_a p]}{p}(@E) \quad \cfrac{}{p \to q}(Axiom)}{q}(\to E)}{@_a q}(@I)}{@_a q}(Term)$$

---
[5]The author thanks Michiel van Lambalgen for mentioning the Smarties task in an email exchange where the author suggested that the shift of perspective in the hybrid-logical rule (*Term*) could be of relevance in connection with the theory of mind view of autism.

**Figure 3: First example formalization (before and after application of the (*Term*) rule)**

**Figure 4: Formalization of the child's reasoning in the Smarties task (both temporal and person version)**

$$\cfrac{@_a Dp \quad \cfrac{[a] \quad \cfrac{\cfrac{[a] \quad [@_a Dp]}{Dp}(@E) \quad \cfrac{}{Dp \to Bp}(Axiom\ schema)}{Bp}(\to E)}{@_a Bp}(@I)}{@_a Bp}(Term)$$

[23].[6] Then the shift of temporal perspective in the Smarties task can be formalized very directly in Seligman's system as the derivation in Figure 4. Recall that the derivation is meant to formalize each step in Peters's reasoning at the time $t$ where the second question is answered. The premise $@_a Dp$ in the derivation says that Peter at the earlier time $a$ deduced that there were Smarties inside the tube, which he remembers at $t$.

Note that the formalization in Figure 4 does not involve the □ operator, so this operator could have been omitted together with the associated rules (□$I$) and (□$E$) in Figure 1. Since this proof system is complete, the □ operator satisfies logical omniscience. The operators $D$ and $B$ are only taken to satisfy the principle $D\phi \to B\phi$, as mentioned above.

Compare the derivation in Figure 4 to the right-hand-side derivation in Figure 3 in the previous section and note that the structure is exactly the same. Note that what we have done is that we have formalized the logical reasoning taking place when giving the correct answer "Smarties". Note also that the actual content of the tube, namely pencils, is not even mentioned in the formalization, so it is clear from the formalization that the actual content of the tube is not relevant to figure out the correct answer. Accordingly, our formalization does not tell what goes wrong when a child incorrectly answers "Pencils".

## 6. THE SMARTIES TASK (PERSON SHIFT VERSION)

As a stepping stone between the temporal version of the Smarties task we considered in the previous section, and the Sally-Anne task we shall consider in the next section, we in the present section take a look again at the version of the Smarties task described in the introduction. The only difference between the version in the introduction and the version in the previous section is the second question where

   "Before this tube was opened, what did you think was inside?"

obviously gives rise to a temporal shift of perspective, whereas

   "If your mother comes into the room and we show this tube to her, what will she think is inside?"

gives rise to a shift of perspective to another person, namely the imagined mother.

To give a correct answer to the latter of these two questions, the child Peter imagines being the mother coming into the room. Imagining being the mother, Peter reasons that the mother must deduce that there are Smarties inside the tube from the fact that it is a Smarties tube, and from that, she must also come to believe that there are Smarties inside. Therefore, Peter concludes that the mother would believe that there are Smarties inside.

The derivation formalizing this argument is exactly the same as in the temporal case dealt with in previous section, Figure 4, but the symbols are interpreted differently, namely as

   $D$   Deduces that ...
   $B$   Believes that ...
   $p$   There are Smarties inside the tube
   $a$   The imagined mother

So now nominals refer to persons rather than times. Accordingly, the modal operator $B$ now symbolize the belief of the person represented by the point of evaluation, rather than Peter's belief at the time of evaluation, etc. Thus, the premise $@_a Dp$ in the derivation in Figure 4 says that the imagined mother deduces that there are Smarties inside the tube, which the child doing the reasoning takes to be the case since the mother is imagined to be present in the room.

---

[6] Analogous to the question in Footnote 4, it can be asked why we use classical implication in $D\phi \to B\phi$, rather than a form of non-monotonic implication. Again, the answer is that this is an idealization, but we presume that the perspective shift involved in the Smarties task is orthogonal to the issue of non-monotonicity, at least from a logical point of view. In this connection we remark that principle (9.4) in [23] also uses classical implication (the non-monotonicity in the logical analysis of the Smarties task of [23] does not concern principle (9.4), but other principles).

Incidentally, letting points in the Kripke model represent persons is exactly what is done in Arthur Prior's *egocentric logic*, see Section 1.3 in the book [8], in particular pp. 15–16. In egocentric logic the accessibility relation represents the taller-than relation, but this relation is obviously not relevant here.

## 7. THE SALLY-ANNE TASK

In this section we will give a formalization of a somewhat more complicated reasoning task called the Sally-Anne task. The following is one version.

> A child is shown a scene with two doll protagonists, Sally and Anne, having respectively a basket and a box. Sally first places a marble into her basket. Then Sally leaves the scene, and in her absence, the marble is transferred by Anne and hidden in her box. Then Sally returns, and the child is asked: "Where will Sally look for her marble?"

Most children above the age of four correctly responds where Sally must falsely believe the marble to be (in the basket) whereas younger children respond where they know the marble to be (in the box). Again, for autists, the cutoff is higher.

Below we shall formalize the correct response to the task, but before that, we give an informal analysis. Let us call the child Peter again. Let $t_1$ be the time where he answers the question. To answer the question, Peter imagines himself being Sally at an earlier time $t_0$ before she leaves the scene, but after she places the marble in her basket. Imagining being Sally, he reasons as follows: At the time $t_0$ Sally believes that the marble is in the box since she can see it. At the time $t_1$, after she has returned, she deduces that the marble is still in the box as she has no belief to the contrary, and since Sally deduces that the marble is in the box, she must also come to believe it. Therefore, Peter concludes that Sally believes that the marble is in the box.

In our formalization we make use of a tiny fragment of first-order hybrid logic, involving the unary predicate $P(t)$, the binary predicate $t < u$, and the modal operators $S$, $D$ and $B$, but no quantifiers. We make use of the following symbolizations

| | |
|---|---|
| $p(t)$ | The marble is in the basket at the time $t$ |
| $t < u$ | The time $t$ is before the time $u$ |
| $S$ | Sees that ... |
| $D$ | Deduces that ... |
| $B$ | Believes that ... |
| $a$ | The person Sally |

We also make use of the following three principles

$$S\phi \rightarrow B\phi$$
$$D\phi \rightarrow B\phi$$
$$B\phi(t) \wedge t < u \wedge \neg B\neg\phi(u) \rightarrow D\phi(u)$$

The first two are versions of principles (9.2) and (9.4) in the book [23] and the third is similar to principle (9.11) in that book. In order to make the formalization more compact, and also more in the spirit of natural deduction style, we do not take the principles as axiom schemas, but instead we turn them into the following proof-rules.

$$\frac{S\phi}{B\phi}\,(R1) \qquad \frac{D\phi}{B\phi}\,(R2) \qquad \frac{B\phi(t) \quad t < u \quad \neg B\neg\phi(u)}{D\phi(u)}\,(R3)$$

The second and third proof-rule together formalizes a "principle of inertia" saying that a belief is preserved over time, unless there is belief to the contrary.

We liberalize the side-condition on the (Term) rule such that the formulas $\phi_1, \ldots, \phi_n$, and $\psi$ may include formulas on the form $t < u$, since we assume that the truth-values of such formulas are not changed by the perspective shift effected by the rule.

With this machinery in place, the shift of person perspective in the Sally-Anne task can be formalized as the derivation in Figure 5. Recall that this derivation is meant to formalize the child's reasoning at the time $t_1$ where the question is answered. The first premise $@_a Sp(t_0)$ in the derivation says that Sally (the reference the nominal $a$) at the earlier time $t_0$ saw that the marble was in the basket, which the child remembers. The third premise $@_a \neg B\neg p(t_1)$ says that Sally at the time $t_1$ does not believe that the marble is not in the basket, which the child realizes as Sally was absent when the marble was transferred to the box.

Note that the actual position of the marble at the time $t_1$ is irrelevant to figure out the correct response. Note that in the Sally-Anne task there is a shift of person perspective which we deal with in a modal-logical fashion letting points of evaluation stand for persons, like in the person version of the Smarties task in the previous section, but there is also a temporal shift in the Sally-Anne task, from the time $t_0$ to the time $t_1$, which we deal with using first-order machinery.

## 8. DISCUSSION

In the introduction of the present paper we remarked that reasoning in Seligman's system is different from reasoning in the most common proof systems for hybrid logic, and that reasoning in Seligman's system captures well the reasoning in the Smarties and Sally-Anne tasks, in particular the involved shift between different local perspectives.

More can be said about this difference between the proof systems and how local perspectives are (or are not) represented. A truth-bearer is an entity that is either true or false. According to Peter Simons' paper [22], there have historically been two fundamentally opposed views of how truth-bearers have their truth-values.

> One view takes truth to be absolute: a truth-bearer's truth-value (whether truth or falsity) is something it has *simpliciter*, without variation according to place, time, by whom and to whom it is said. The other view allows a truth-bearer's truth-value to vary according to circumstances: typically time or place, but also other factors may be relevant. ([22], p. 443)

Peter Simons calls the first view the *absolute* view and the second the *centred* view. It is well-known that Arthur Prior often expressed sympathy for what is here called the centred view, most outspoken with respect to time, one reason being that he wanted to allow statements to change truth-value from one time to another. What a truth-bearer's truth-value varies according to, is by Simons called a *location*.

> I understand 'location' broadly to include not just spatial location but also temporal location, spatiotemporal location, modal location, and more broadly still location in any relational structure. I consider that the concept of an object being

192

Figure 5: Formalization of the child's reasoning in the Sally-Anne task

$$\cfrac{@_a Sp(t_0) \quad t_0 < t_1 \quad @_a \neg B \neg p(t_1) \quad \cfrac{[a] \quad \cfrac{\cfrac{[a] \quad [@_a Sp(t_0)]}{Sp(t_0)}(@E)}{Bp(t_0)}(R1) \quad [t_0 < t_1] \quad \cfrac{[a] \quad [@_a \neg B \neg p(t_1)]}{\neg B \neg p(t_1)}(@E)}{\cfrac{Dp(t_1)}{Bp(t_1)}(R2)}(R3)}{@_a Bp(t_1)}(Term)}{@_a Bp(t_1)}$$

located at a position among other positions is a formal concept, applicable topic-neutrally in any field of discourse. This means that logical considerations about location are not limited in extent or parochial in interest. ([22], p. 444)

The proposition expressed in the quotation above is defended in Simons' paper [21]. See also the paper [20]. Obviously, a frame for modal and hybrid logic is a mathematically precise formulation of Simons' concept of a location, see Definition 2.1.

What does all this have to do with proof systems for hybrid logic? The distinction between the absolute view and the centred view is useful for describing proof systems and the formulas that occur in them. The basic building blocks of the most common proof systems for hybrid logic are satisfaction statements, and satisfaction statements have constant truth-values, so the basic building blocks of such systems are absolute, although it is arguable that such systems have both absolute and centred features since arbitrary subformulas of satisfaction statements do have varying truth-values, and therefore have to be evaluated for truth at some location. On the other hand, the basic building blocks of Seligman's system are arbitrary formulas, and arbitrary formulas have varying truth-values, so this system is centred, involving local perspectives in the reasoning.

## 9. SOME REMARKS ON OTHER WORK

Beside analysing the reasoning taking place when giving a correct answer to a reasoning task, the works by van Lambalgen and co-authors also analyse what goes wrong when an incorrect answer is given. We note that Stenning and van Lambalgen in [23] warn against simply characterizing autism as a lack of theory of mind. Rather than being an explanation of autism, Stenning and van Lambalgen see the theory of mind deficit hypothesis as "an important label for a problem that needs a label", cf. [23], p. 243. Based on their logical analysis, they argue that another psychological theory of autism is more fundamental, namely what is called the *executive function deficit theory*. Very briefly, executive function is an ability to plan and control a sequence of actions with the aim of obtaining a goal in different circumstances.

The paper [15] reports empirical investigations of closed-world reasoning in adults with autism. Incidentally, according to the opening sentence of that paper, published in 2009, "While autism is one of the most intensively researched psychiatric disorders, little is known about reasoning skills of people with autism."

With motivations from the theory of mind literature, the paper [25] models examples of beliefs that agents may have about other agents' beliefs (one example is an autistic agent that always believes that other agents have the same beliefs as the agent's own). This is modelled by different agents preference relations between states, where an agent prefers one state over another if the agent considers it more likely. The beliefs in question turn out to be frame-characterizable by formulas of epistemic logic.

The paper [10] reports empirical investigations of what is called *second-order* theory of mind, which is a person's capacity to imagine other people's beliefs about the person's own beliefs (where *first-order* theory of mind is what we previously in the present paper just have called theory of mind). The investigations in [10] make use of a second-order false-belief task, as well as other tasks.

The paper [12] does not deal with false-belief tasks or theory of mind, but it is nevertheless relevant to mention since it uses formal proofs to compare the cognitive difficulty of deductive tasks. To be more precise, the paper associates the difficulty of a deductive task in a version of the Mastermind game with the minimal size of a corresponding tableau tree, and it uses this measure of difficulty to predict the empirical difficulty of game-plays, for example the number of steps actually needed for solving a task.

The method of reasoning in tableau systems can be seen as attempts to construct a model of a formula: A tableau tree is built step by step using rules, whereby more and more information about models for the formula is obtained, and either at some stage a model can be read off from the tableau tree, or it can be concluded that there cannot be such a model (in fact, in the case of [12], any formula under consideration has exactly one model, so in that case it is a matter of building a tableau tree that generates this model). Hence, if the building of tableau trees is taken to be the underlying mechanism when a human is solving Mastermind tasks, then the investigations in [12] can be seen to be in line with the mental models school (see the third section of the present paper).

A remark from a more formal point of view: The tableau system described in [12] does not include the cut-rule[7]. Much has been written on the size of proofs in cut-free proof systems, in particular, the paper [6] gives examples of first-order formulas whose derivations in cut-free systems are much larger than their derivations in natural deduction systems, which implicitly allow unrestricted cuts (in one case more than $10^{38}$ characters compared to less than 3280 characters). Similarly, the paper [9] points out that ordinary cut-free tableau systems have a number of anomalies, one of them being that for some classes of propositional formulas, decision procedures based on cut-free systems are much slower than the truth-table method (in the technical sense that there is no polynomial time computable function that maps truth-table proofs of such formulas to proofs of the same formulas in cut-free tableau systems). Instead of prohibiting cuts completely, the paper [9] advocates allowing a restricted version of the cut-rule, called the analytic cut-rule.

---

[7]The cut-rule says that the end of any branch in a tableau tree can extended with two branches with $\phi$ on the one branch and $\neg \phi$ on the other (expressing the bivalence of classical logic).

## 10. FUTURE WORK

We would like to extend the work of the present paper to further false-belief tasks, perhaps using different hybrid-logical machinery (and moreover, to see if we can also use hybrid-logical proof-theory to analyse what goes wrong when incorrect answers are given). Not only will formalization of further reasoning tasks be of interest on their own, but we also expect that such investigations can be feed back into logical research, either as corroboration of the applicability of existing logical constructs, or in the form of new logical constructs, for example new proof-rules or new ways to add expressive power to a logic.

We are also interested in further investigations in when two seemingly dissimilar reasoning tasks have the same underlying logical structure, like we in the present paper have disclosed that two different versions of the Smarties task have exactly the same underlying logical structure. Such investigations might be assisted by a notion of identity on proofs (exploiting the longstanding effort in proof-theory to give a notion of identity between proofs, that is, a way to determine if two arguments have common logical structure, despite superficial dissimilarity).

More speculatively, we expect that our formalizations can contribute to the ongoing debate between two dominating views on theory of mind, denoted *theory-theory* and *simulation-theory*. According to theory-theory, theory of mind should be viewed as an explicit theory of the mental realm of another person, like the theories of the physical world usually going under the heading "naive physics", whereas according to simulation-theory, theory of mind should be viewed as a capacity to put yourself in another person's shoes, and simulate the person's mental states.

## 11. ACKNOWLEDGEMENTS

The author thanks Thomas Bolander for comments on an early version of this paper. Also thanks to Jerry Seligman for a discussion of the paper. The author acknowledges the financial support received from The Danish Natural Science Research Council as funding for the project HYLOCORE (Hybrid Logic, Computation, and Reasoning Methods, 2009–2013).

# APPENDIX
## A. PROOF OF ANALYTICITY

Usually, when considering a natural deduction system, one wants to equip it with a normalizing set of reduction rules such that normal derivations satisfy the subformula property. Normalization says that any derivation by repeated applications of reduction rules can be rewritten to a derivation which is normal, that is, no reduction rules apply. From this it follows that the system under consideration is analytic.

Now, the works [7] and Section 4.3 by the present author devise a set of reduction rules for $\mathbf{N}'_\mathcal{H}$ obtained by translation of a set of reduction rules for a more common natural deduction system for hybrid logic. This more common system, which we denote $\mathbf{N}_\mathcal{H}$, can be found in [7] and in [8], Section 2.2. All formulas in the system $\mathbf{N}_\mathcal{H}$ are satisfaction statements. Despite other desirable features, it is not known whether the reduction rules for $\mathbf{N}'_\mathcal{H}$ are normalizing, and normal derivations do not always satisfy the subformula property. In fact, Chapter 4 of the book [8] ends somewhat pessimistically by exhibiting a normal derivation without the subformula property. It is remarked that a remedy would be to find a more complete set of reduction rules, but the counter-example does not give a clue how such a set of reduction rules should look.

In what follows we shall take another route. We prove a completeness result saying that any valid formula has a derivation in $\mathbf{N}'_\mathcal{H}$ satisfying a version of the subformula property. This is a sharpened version of a completeness result for $\mathbf{N}'_\mathcal{H}$ originally given in [7] and in Section 4.3 of [8] (Theorem 4.1 in [8]). Thus, we prove that $\mathbf{N}'_\mathcal{H}$ is analytic without going via a normalization result. So the proof of the completeness result does not involve reduction rules. The result is mathematically weaker than normalization together with the subformula property for normal derivations, but it nevertheless demonstrates analyticity. Analyticity is a major success criteria in proof-theory, one reason being that analytic provability is a step towards automated theorem proving (which obviously is related to Leibniz' aim mentioned in the intoduction of the present paper).

In the proof below we shall refer to $\mathbf{N}_\mathcal{H}$ as well as a translation $(\cdot)^\circ$ from $\mathbf{N}_\mathcal{H}$ to $\mathbf{N}'_\mathcal{H}$ given in [7] and Section 4.3 of [8]. This translates a derivation $\pi$ in $\mathbf{N}_\mathcal{H}$ to a derivation $\pi^\circ$ in $\mathbf{N}'_\mathcal{H}$ having the same end-formula and parcels of undischarged assumptions. The reader wanting to follow the details of our proof is advised to obtain a copy of the paper [7] or the book [8]. The translation $(\cdot)^\circ$ satisfies the following.

LEMMA A.1. *Let $\pi$ be a derivation in $\mathbf{N}_\mathcal{H}$. Any formula $\theta$ occuring in $\pi^\circ$ has at least one of the following properties.*

1. *$\theta$ occurs in $\pi$.*
2. *$@_a\theta$ occurs in $\pi$ for some satisfaction operator $@_a$.*
3. *$\theta$ is a nominal $a$ such that some formula $@_a\psi$ occurs in $\pi$.*

PROOF. Induction on the structure of the derivation of $\pi$. Each case in the translation $(\cdot)^\circ$ is checked. □

Note that in item 1 of the lemma above, the formula $\theta$ must be a satisfaction statement since only satisfaction statements occur in $\pi$. In what follows $@_d\Gamma$ denotes the set of formulas $\{@_d\xi \mid \xi \in \Gamma\}$.

THEOREM A.2. *Let $\pi$ be a normal derivation of $@_d\phi$ from $@_d\Gamma$ in $\mathbf{N}_\mathcal{H}$. Any formula $\theta$ occuring in $\pi^\circ$ has at least one of the following properties.*

1. *$\theta$ is of the form $@_a\psi$ such that $\psi$ is a subformula of $\phi$, some formula in $\Gamma$, or some formula of the form $c$ or $\Diamond c$.*
2. *$\theta$ is a subformula of $\phi$, some formula in $\Gamma$, or some formula of the form $c$ or $\Diamond c$.*
3. *$\theta$ is a nominal.*
4. *$\theta$ is of the form $@_a(p \to \bot)$ or $p \to \bot$ where $p$ is a subformula of $\phi$ or some formula in $\Gamma$.*
5. *$\theta$ is of the form $@_a\bot$ or $\bot$.*

PROOF. Follows from Lemma A.1 above together with Theorem 2.4 (called the quasi-subformula property) in Subsection 2.2.5 of [8]. □

We are now ready to give our main result, which is a sharpened version of the completeness result given in Theorem 4.1 in Section 4.3 of [8].

THEOREM A.3. *The first statement below implies the second statement. Let $\phi$ be a formula and $\Gamma$ a set of formulas.*

1. *For any model $\mathcal{M}$, any world $w$, and any assignment $g$, if, for any formula $\xi \in \Gamma$, $\mathcal{M}, g, w \models \xi$, then $\mathcal{M}, g, w \models \phi$.*
2. *There exists of derivation of $\phi$ from $\Gamma$ in $\mathbf{N}'_\mathcal{H}$ such that any formula $\theta$ occuring in the derivation has at least one of the five properties listed in Theorem A.2.*

PROOF. Let $d$ be a new nominal. It follows that for any model $\mathcal{M}$ and any assignment $g$, if, for any formula $@_d\xi \in @_d\Gamma$, $\mathcal{M}, g \models @_d\xi$, then $\mathcal{M}, g \models @_d\phi$. By completeness of the system $\mathbf{N}_\mathcal{H}$, Theorem 2.2 in Subsection 2.2.3 of the book [7], there exists a derivation $\pi$ of $@_d\phi$ from $@_d\Gamma$ in $\mathbf{N}_\mathcal{H}$. By normalization, Theorem 2.3 in Subsection 2.2.5 of the book, we can assume that $\pi$ is normal. We now apply the rules $(@I)$, $(@E)$, and $(Name)$ to $\pi^\circ$ obtaining a derivation of $\phi$ from $\Gamma$ in $\mathbf{N}'_\mathcal{H}$ satisfying at least one of the properties mentioned in Theorem A.2. □

Remark: If the formula occurrence $\theta$ mentioned in the theorem above is not of one of the forms covered by item 4 in Theorem A.2, and does not have one of a finite number of very simple forms not involving propositional symbols, then either $\theta$ is a subformula of $\phi$ or some formula in $\Gamma$, or $\theta$ is of the form $@_a\psi$ such that $\psi$ is a subformula of $\phi$ or some formula in $\Gamma$. This is the version of the subformula property we intended to prove.